\documentclass[]{siamart1116}

\usepackage{amsmath}
\usepackage{makeidx}
\usepackage{amsfonts}
\usepackage{amssymb}
\usepackage[utf8]{inputenc}
\usepackage[usenames,dvipsnames]{pstricks}
\usepackage{subfigure}
\usepackage{epsfig}
\usepackage{pst-grad} 
\usepackage{pst-plot} 
\usepackage{placeins}
\usepackage{tabto}
\usepackage{bm}
\usepackage{dsfont}

\numberwithin{theorem}{section}

\newcommand{\TheTitle}{Control of Nonlinear Wave Solutions to Neural Field Equations} 
\newcommand{\TheAuthors}{A. Ziepke, S. Martens, and H. Engel}

\headers{\TheTitle}{\TheAuthors}

\title{{\TheTitle}
}

\author{
  Alexander Ziepke\thanks{Technische Universit\"at Berlin, EW 7-1, Institut f\"ur Theoretische Physik, Hardenbergstra\ss e 36, 10623 Berlin, Germany
    (\email{ziepke@itp.tu-berlin.de}).}
  \and
  Steffen Martens\footnotemark[1]
  \and
  Harald Engel\footnotemark[1]
}

\usepackage{amsopn}

\ifpdf
\hypersetup{
  pdftitle={\TheTitle},
  pdfauthor={\TheAuthors}
}
\fi

\begin{document}

\maketitle

\begin{abstract}
Neural field equations offer a continuous description of the dynamics of large populations of synaptically coupled neurons. 
This makes them a convenient tool to describe various neural processes, such as working memory, motion perception, and visual hallucinations, to name a few. Due to the important applications, the question arises how to effectively control solutions in such systems.\\
In this work, we investigate the problem of position control of traveling wave solutions to scalar neural field equations on the basis of singular perturbation analysis. Thereby, we consider different means of control such as spatio-temporal modulations of the neural firing threshold, asymmetric synaptic coupling kernels, and additive inputs. Treating these controls as perturbations to the neural field system, one obtains an equation of motion for traveling wave solutions in response to the applied controls. Subsequently, we pose the inverse question of how to design controls that lead to a propagation of the solution following a desired velocity protocol.
In particular, we make use of a specific excitation of the solution's translational modes \cite{Lober2014} which enables an explicit calculation of a necessary control signal for a given velocity protocol without evoking shape deformations. Moreover, we derive an equivalent control method relying on a modulation of the neurons' synaptic footprint.
\end{abstract}



\section{Introduction}
Neural field equations serve as an important instrument for modeling the dynamics of a large number of synaptically coupled neurons by means of continuous field equations \cite{Coombes2005}. Based on seminal works by Wilson and Cowan \cite{Wilson1972} and Amari \cite{Amari1977}, the theory for such populations of coupled neurons evolved and nowadays is broadly accepted as a model system for various neuro-biological effects such as the functionality of working memory \cite{Kilpatrick2013a,Laing2003b,Lins2014}, visual hallucinations \cite{{Coombes2012},{Ermentrout1979}}, and detection of spatial orientation \cite{{Taube2003},{Zhang1996}}.
In particular, many applications make use of the rich variety of solutions to the neural field equations (NFEs). For example, traveling wave (TW) solutions like fronts and pulses \cite{{Atay2005},{Pinto2001a}}, spiral waves \cite{Laing2005}, and spot-like bump solutions \cite{{Coombes2005a},{Lu2011}} are of special interest. They are related to regions of higher neural activity and therefore a fundamental model object for information processing in neural cortex \cite{Bressloff2014Neural}.\\
Further applications of neural fields (NFs) can be found in biologically motivated robotics. Here, in analogy to human information processing, the equations can be used to model motor responses to external stimuli as well as change recognition \cite{Erlhagen2006}. Within this framework, there are approaches controlling the position of robot arms by means of shifting the spatial position of a bump solution in a corresponding abstract space \cite{Fard2015}.
Not only for such applications a systematic discussion of position control is desirable.
For instance, the treatment of neurological disorders such as Alzheimer's and Parkinson's disease, migraine, and epilepsy still demands for a deeper understanding of brain functioning. Therefore, insights on wave propagation in neural tissue and mechanisms to control activity regions are relevant \cite{Goadsby2002,schwartzkroin2007epilepsy,Tass2007}.\\
As a first step in this direction, the response of TW solutions in NFEs to external stimuli attracted interest of several groups over the past few years. For instance, homogenization methods on the basis of an asymptotic perturbation analysis have been used to describe wave propagation in neural fields with spatially modulated synaptic weights \cite{Bressloff2001,Coombes2011,Kilpatrick2008}. Similar methods have been applied for investigation of the response of TW solutions to perturbations by external additive inputs \cite{Ermentrout2010,Kilpatrick2012}, fluctuations \cite{Kilpatrick2015,Kilpatrick2013}, or for predicting the propagation of multiple bump solutions interacting within periodic pulse trains \cite{Bressloff2005}.
Besides these efforts in understanding the response of TW solutions to perturbations, approaches have been made to control stationary solutions by means of asymmetric coupling functions \cite{Zhang1996} and external inputs \cite{Lu2011}.
Additionally, inverse problems in neural field equations were treated by Potthast and beim Graben applying Hebbian learning techniques to discretized neural field systems \cite{Potthast2009a}. 
Furthermore, there are experiments on the influence of external electric fields on pulse transmission in neural cortex.
For example, an applied electric field is used to modulate the firing threshold of aligned neurons and, thus, enables a control of signal transmission velocities \cite{Richardson2005}.
Moreover, optogenetics is a promising technique making a direct control of neural firing activity experimentally accessible \cite{Deisseroth2011,Fenno2011,Grosenick2015}.\\
With respect to the important applications of neural field equations, we systematically treat position control on the basis of singular perturbation theory \cite{Segel1989,Tyson1988}. 
In particular, we examine the implementation of open-loop control that forces stable TW solutions in one spatial dimension to move according to a pre-defined protocol of motion. This method solely bases on preliminary measurements and renders monitoring of the progress unnecessary.
%
Moreover, we investigate the stability of the resulting schemes with respect to emerging offsets.\\
After reviewing the standard scalar neural field equation in \cref{sec:NFEqs}, we state the problem and derive a scheme for position control of TW solutions in \cref{sec:Control}. Following this, we explicitly treat position control by additive inputs, spatio-temporal modulations of the neural firing threshold as well as asymmetries in the synaptic footprint in \cref{sec:Control_app}. In particular, we make use of a direct excitation of the solutions translational mode by an external input and derive an equivalent control by kernel modulations.
Finally, we conclude our analysis in \cref{sec:Conclusion}.

\section{Traveling Wave Solutions to the Neural Field Equations}
\label{sec:NFEqs}
\begin{figure}[!tb]
\center
\includegraphics[width=0.75\textwidth]{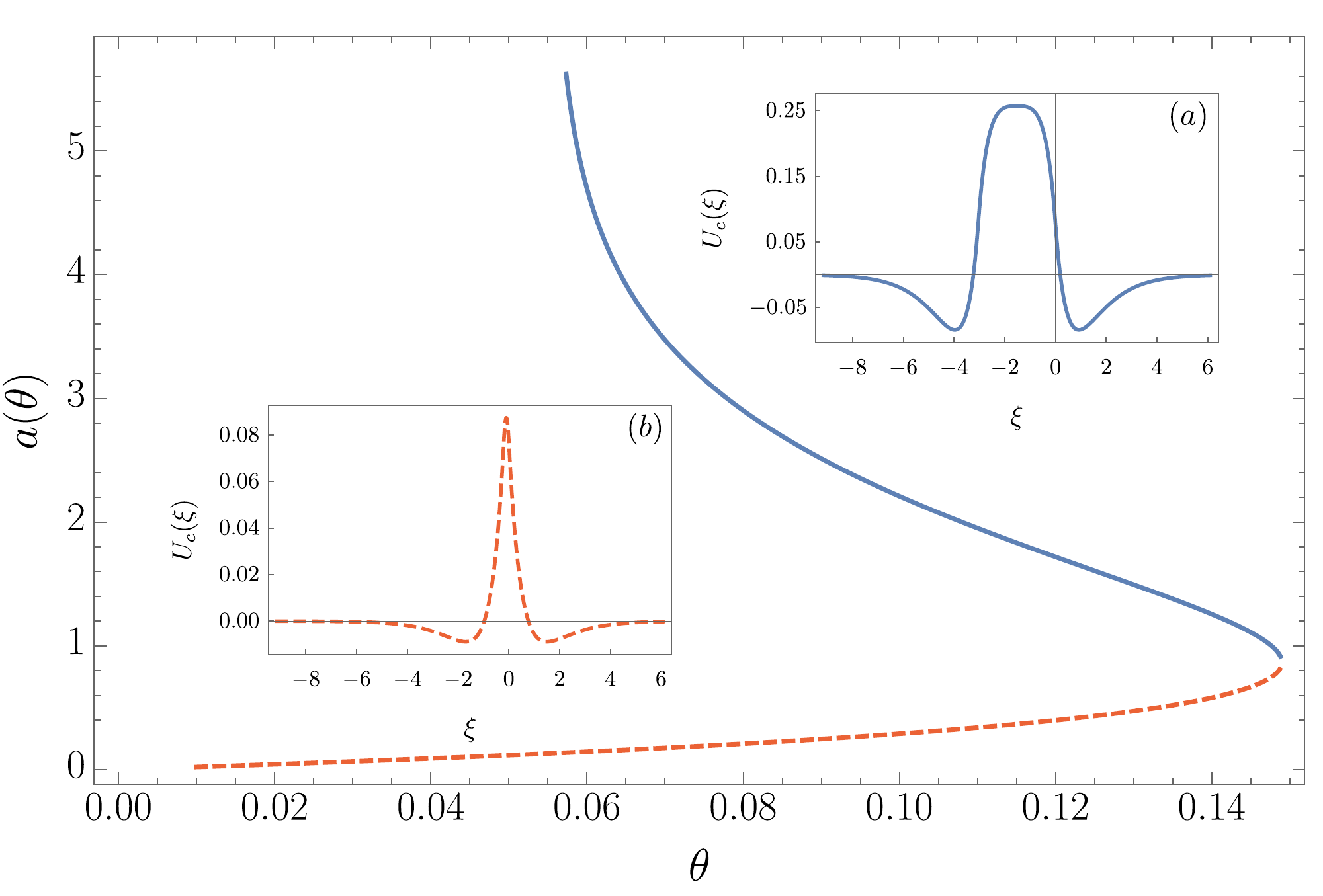}
\caption{\label{fig:solBump}Immobile bump solutions to the scalar neural field equation, \cref{eq:Neural_Field}. The two branches of solutions with threshold dependent bump width $a(\theta)$ are shown for a spatial coupling kernel $\omega(x)=e^{-1.8\lvert x\rvert}-0.5e^{-\lvert x\rvert}$. The solutions of the broader branch (solid blue line) are stable, whereas the other branch (dashed orange line) represents narrower unstable bumps. The insets $(a)$ and $(b)$ show corresponding profiles for $\theta=0.077$.}
\end{figure}

Throughout this paper, we restrict our considerations to scalar neural fields \cite{Bressloff2011} in one spatial dimension, $x\in\mathbb{R}$,
\begin{align}\label{eq:Neural_Field}
\partial_t u(x,t)=-u(x,t)+\int_{-\infty}^{\infty}\omega(x-y)f(u(y,t))\mathrm{d}y.
\end{align}
The dynamics of the activity variable $u(x,t)$ is governed by a local exponential decay with unit timescale and a non-local input being modeled by a spatial convolution of a nonlinear firing function $f(u)$ with the synaptic coupling kernel $\omega(x)$.
Following approximations for the high gain limit of neural firing \cite{Amari1977}, we consider a neural field system with Heaviside nonlinearity 
\begin{align}
f\left(u(x,t)\right)=\Theta(u(x,t)-\theta)=\begin{cases}
1 &,u(x,t) > \theta,\\
0 &\mbox{, else,}
\end{cases}
\end{align}
and threshold parameter $\theta\in\mathbb{R}$. 
The scalar neural field equation \cref{eq:Neural_Field} exhibits various traveling wave solutions such as fronts for exponentially decaying excitatory synaptic coupling and immobile bumps for kernels with local excitation and lateral inhibition \cite{Amari1977,Coombes2005}.
Transforming space into the co-moving wave coordinate $\xi=x-c\,t$ with propagation velocity $c$ in \cref{eq:Neural_Field}, one obtains the time independent ordinary differential equation for stationary TW solutions $U_c(\xi)$,
\begin{align}\label{eq:NFE_TW}
\left(1-c\partial_{\xi}\right)U_c(\xi)=\int_{-\infty}^{\infty}\omega(\xi-\xi')\Theta\left(U_c(\xi')-\theta\right)\mathrm{d}\xi'.
\end{align} 
For monotonous front solutions, we chose the single threshold crossing point to be located at $\xi=0$, yielding, $U_c(\xi)>\theta\,\forall\,\xi<0$ and $U_c(\xi)<\theta\,\forall\,\xi>0$. With the integrating factor $-e^{-\xi/c}$, one gets
\begin{align}\label{eq:front_int}
U_c(\xi)=e^{\xi/c}\left(\theta-\frac{1}{c}\int_0^{\xi}e^{-\xi'/c}\int_{\xi'}^{\infty}\omega(\xi'')\mathrm{d}\xi''\mathrm{d}\xi'\right).
\end{align}
Thus, the traveling front solution to \cref{eq:NFE_TW} with the exponentially decaying coupling kernel $\omega(x)=0.5e^{-\lvert x\rvert}$ reads \cite{Ermentrout1993}
\begin{align}\label{eq:front}
U_c(\xi)=\begin{cases}
\theta e^{-\xi},&\xi\geq 0,\\
\frac{\theta}{1-4\theta}e^{\xi}+1-\frac{(1-2\theta)^2}{1-4\theta}e^{2\theta/(1-2\theta)\xi}, &\xi < 0.
\end{cases}
\end{align}
Its propagation velocity can be calculated to be $c=(1-2\theta)/(2\theta)$, requiring boundedness of $U_c(\xi)$ for $\xi\rightarrow\infty$, cf.\@ \cref{eq:front_int}.
\begin{figure}[!tb]
\center
\includegraphics[width=0.6\textwidth]{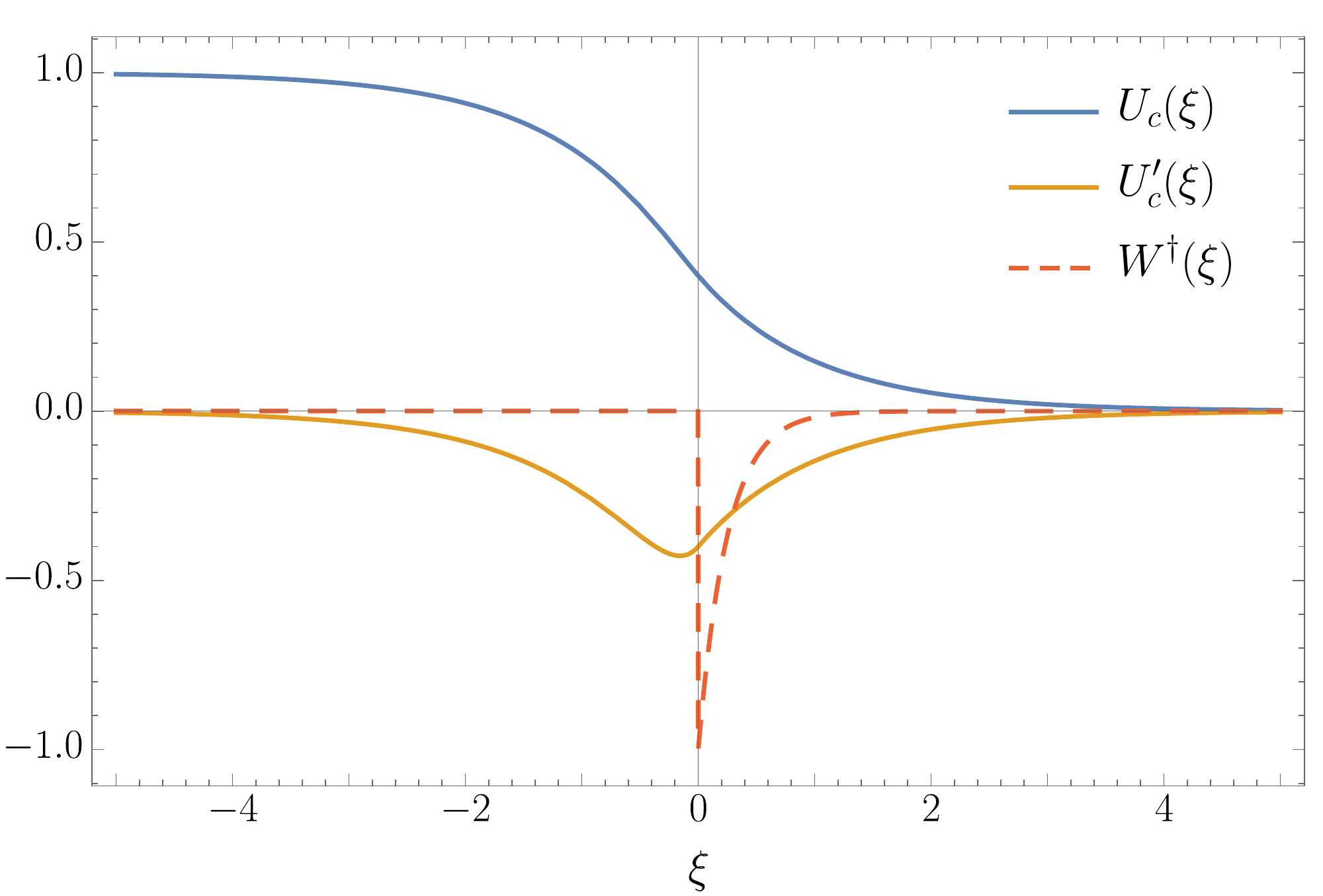}
\caption{\label{fig:solFront}Traveling front solution $U_c(\xi)$, see \cref{eq:front}, to \cref{eq:NFE_TW}, its first spatial derivative $U_c'(\xi)$, and response function $W^{\dagger}(\xi)$, see \cref{sec:Control}. The synaptic coupling function is set to $\omega(x)=0.5e^{-\lvert x\rvert}$ and the threshold of the Heaviside firing function is $\theta=0.4$.}
\end{figure}
The front solution, depicted in \cref{fig:solFront}, resembles a bistable configuration of neural activity, viz., a transition from a quiescent to an active firing state or vice versa. 
In real neural systems, intrinsic inhibitory mechanisms play an important role as they lead to a reduction of neuronal activity after some time of persistent firing \cite{Izhikevich2007Dynamical}. Nevertheless, the front solution can be seen as an approximation for the interface of a broadening or propagating localized region of increased neural activity. \\
%
Incorporating inhibitory neural synapses or local mechanisms like synaptic depression or spike frequency adaptation that can suppress the neuron's activity, the neural field equation \cref{eq:Neural_Field} additionally exhibits bump solutions which represent such localized regions of neural activity \cite{Laing2002a}. For stationary immobile bump solutions, $\partial_t u(x,t)=0$, equation \cref{eq:Neural_Field} can be solved analytically demanding that the activity is above threshold only within a finite domain of width $a$ and below anywhere else on the real axis, viz., $U_c(\xi)>\theta\,\forall\,\xi\in (-a,0)$ and $U_c(\xi)<\theta\,\forall\,\xi\in\mathbb{R}\setminus\left[-a,0\right]$. By virtue of the Heaviside firing function, one gets
\begin{align}\label{eq:bump_sol}
U_c(\xi)=\int_{\xi}^{\xi+a}\omega(y)\mathrm{d}y,
\end{align}
with bump width $a$. The latter can be derived from the threshold condition
\begin{align}
\theta=U_c(-a)=\int_{-a}^{0}\omega(y)\mathrm{d}y.
\end{align}
In \cref{fig:solBump}, the two branches of immobile bump solutions are shown for the ``mexican-hat'' coupling function 
\begin{align}\label{eq:kernel_bump}
\omega(x)=e^{-1.8\lvert x\rvert}-0.5e^{-\lvert x\rvert}.
\end{align}
On the basis of Evans functions, the linear stability of the corresponding solutions can be investigated \cite{Coombes2004}. It turns out that the branch of broader solutions (blue solid line) represents linearly stable bumps whereas the narrower solutions of the other branch (orange dashed line) are unstable. Exemplarily, a stable and an unstable solution for $\theta=0.077$ are depicted in insets $(a)$ and $(b)$ of \cref{fig:solBump}, respectively.
For position control we restrict our considerations to stable TW solutions, as these are supposed to be exhibited by real neural systems. Additionally, such solutions are stable with respect to sufficiently small control signals, and thus are likely to fulfill the assumption of shape conservation throughout the course of control.

%
%

\section{Position Control of Traveling Wave Solutions}\label{sec:Control}

To motivate and derive a general scheme for position control, we first consider the influence of small perturbations on the propagation of TW solutions to the neural field equation \cref{eq:Neural_Field}. We perform a singular perturbation analysis, which is well known for reaction-diffusion systems \cite{Lober2014,Martens2015a,Tyson1988,Ziepke2016} and was similarly applied to neural fields \cite{Bressloff2005,Ermentrout2010,Kilpatrick2008}.
As we aim to control the spatial position of stable TW solutions, we are particularly interested in their translational response to applied perturbations. We consider different kinds of controls, namely, additional additive inputs $\epsilon I(x,t)$ \cite{Kilpatrick2012,Lu2011}, modified coupling kernels $\omega(x)=\omega_0(x)+\epsilon\omega_1(x)$ \cite{Bick2015,Zhang1996}, and spatio-temporal modulations of the firing threshold $\theta(x,t)=\theta_0+\epsilon\theta_1(x,t)$ \cite{Richardson2005}.
All these different perturbations can lead to a break of symmetry and, as a consequence, change the solution's propagation velocity.\\
Incorporating the mentioned perturbations, the NFE reads
\begin{align}\label{eq:perturbedNF}
\partial_tu(x,t)=-u(x,t)+\int_{-\infty}^{\infty}&\left[\omega_0(x-y)+\epsilon\omega_1(x-y)\right]\\ &\times \Theta\left(u(y,t)-(\theta_0+\epsilon\theta_1(y,t))\right)\mathrm{d}y+\epsilon I(x,t).\nonumber
\end{align}
Assuming that perturbations remain small, $\epsilon\ll1$, we expand the solution of the perturbed system about the free solution to \cref{eq:NFE_TW}, $u(\xi,t)=U_c(\xi)+\epsilon u_1(\xi,t)+\mathcal{O}(\epsilon^2)$, with new space coordinate $\xi=x-\phi(t)$, co-moving with the perturbed solution. Additionally, the propagation velocity of the TW is supposed to differ from the unperturbed velocity only by small corrections, viz., $\dot{\phi}(t)=c+\epsilon\dot{\phi}_1(t)+\mathcal{O}(\epsilon^2)$, where a dot denotes the first derivative with respect to time $t$.
Inserting these ansatzes into equation \cref{eq:perturbedNF}, expanding in $\epsilon$, and gathering first order contributions $\mathcal{O}(\epsilon)$, we obtain
\begin{align}
\partial_{t}u_1(\xi,t)=&-u_1(\xi,t)+c\partial_{\xi}u_1(\xi,t)+\dot{\phi}_1(t)\partial_{\xi}U_c(\xi)+I(\xi+\phi(t),t)\\
&+\int_{-\infty}^{\infty}\omega_0(\xi-y)\delta\left(U_c(y)-\theta_0\right)\left[u_1(y,t)-\theta_1(y+\phi(t),t)\right]\mathrm{d}y\nonumber\\
&+\int_{-\infty}^{\infty}\omega_1(\xi-y)\Theta(U_c(y)-\theta_0)\mathrm{d}y.\nonumber
\end{align}
This equation can be reformulated as the inhomogeneous linear operator equation 
\begin{align}\label{eq:linOp}
\partial_{t}u_1(\xi,t)-\mathcal{L}u_1(\xi,t)=&\,\dot{\phi}_1(t)\partial_{\xi}U_c(\xi)\\
&-\int_{-\infty}^{\infty}\omega_0(\xi-y)\delta(U_c(\xi)-\theta_0)\theta_1(y+\phi(t),t)\mathrm{d}y\nonumber\\&+I(\xi+\phi(t),t)+\int_{-\infty}^{\infty}\omega_1(\xi-y)\Theta(U_c(y)-\theta_0)\mathrm{d}y\nonumber,
\end{align}
with linear stability operator $\mathcal{L}=-\mathds{1}+c\partial_{\xi}+\int_{-\infty}^{\infty}\mathrm{d}y\,\omega_0(\xi-y)\delta(U_c(y)-\theta_0)\times$.
In order to derive an equation of motion (EOM) for the perturbed TW solution, we use the Fredholm alternative \cite{Keener1988}. It states that there exists a bounded solution to an inhomogeneous linear operator equation $\mathcal{L}u=f$ if and only if the equation's right-hand side's projection to the nullspace of the adjoint operator $\mathcal{L}^{\dagger}$ is zero.
To apply this solvability condition, we first multiply \cref{eq:linOp} with the vector $W^{\dagger}$ from the nullspace of the adjoint operator using the function space inner product $\langle a(x),b(x)\rangle=\int\mathrm{d}x a^{\ast}(x)b(x)$. The asterisk indicates complex conjugation.
Subsequently, we demand that the wave's shape deformations do not contribute to any displacements. With the stationary response function $W^{\dagger}(\xi)$, one obtains $\langle W^{\dagger},\partial_{t}u_1\rangle=\partial_{t}\langle W^{\dagger},u_1\rangle=0$. Using the solvability condition $\langle W^{\dagger},\mathcal{L}u_1\rangle=0$, we finally end up with
\begin{align}
\label{eq:EOM}
\dot{\phi}_1(t)=-\frac{1}{K_c}\int_{-\infty}^{\infty}W^{\dagger}(\xi) &\left[I(\xi+\phi(t),t)+\int_{-\infty}^{\infty}\omega_1(\xi-\xi')\Theta(U_c(\xi')-\theta_0)\mathrm{d}\xi'\right.\\
&\left. -\int_{-\infty}^{\infty}\omega_0(\xi-\xi')\delta(U_c(\xi')-\theta_0)\theta_1(\xi'+\phi(t),t)\mathrm{d}\xi'\right]\mathrm{d}\xi,\nonumber
\end{align}
with the normalization constant
\begin{align}\label{eq:kc}
K_c=\langle W^{\dagger}(\xi),U_c'(\xi)\rangle=\int_{-\infty}^{\infty}W^{\dagger}(\xi)U_c'(\xi)\mathrm{d}\xi .
\end{align}
Here, a prime denotes the spatial derivative.
Equation \cref{eq:EOM} serves as an EOM for TWs under the considered perturbations. It enables us to calculate the resulting change $\dot{\phi}_1(t)$ of the TW solution's propagation velocity in response to perturbations.\\
In a next step, we regard the EOM from a different point of view. Namely, we pose the inverse question: Which perturbations lead to a propagation of the TW solution in accordance with a desired velocity protocol $\dot{\phi}(t)$. For this purpose, we solve the EOM \cref{eq:EOM} for the perturbations $I(\xi,t)$, $\omega_1(\xi,t)$, and $\theta_1(\xi,t)$ which will be used as means of control. In general, the exact inversion of \cref{eq:EOM} for the control amplitudes requires a knowledge of the solution's response function $W^{\dagger}(\xi)$. Hence, it is possible for only a small number of simple but nevertheless important model systems and controls.\\ 
An important exception to this restriction is the \textit{Goldstone control} which has recently been suggested and applied for position control of solitary solutions to reaction-diffusion systems \cite{Lober2014,Martens2017,Ryll2016}.
The scheme relies on a targeted excitation of the solution's translational modes, the \textit{Goldstone modes}, by an external input. These modes are elements of the nullspace of the linear stability operator $\mathcal{L}$.
Using the ansatz $I(x,t)=\kappa _{\mathrm{Gold}}(t)U_c'(x-\phi(t))$, the EOM \cref{eq:EOM} can always be explicitly inverted for the control amplitude $\kappa_{\mathrm{Gold}}(t)$,
\begin{align}
\dot{\phi}_1(t)&=-\frac{\kappa_{\mathrm{Gold}}(t)}{K_c}\int_{-\infty}^{\infty}W^{\dagger}(\xi)U_c'(\xi)\mathrm{d}\xi, \nonumber\\
\mbox{with}\quad \kappa_{\mathrm{Gold}}(t)&=-\dot{\phi}_1(t).
\end{align}
We emphasize that this renders a knowledge of the solution's response function unnecessary for computing an equation for the control amplitude. All the information about the system that is necessary to set up the Goldstone control is encoded within the first spatial derivative of the unperturbed solution, e.g., the uncontrolled wave profile, and its propagation velocity. Hence, the technique offers the possibility to calculate the control signal for a pre-defined velocity protocol in advance for any kind of stable stationary solution.
To illustrate the generality of the derived control ansatz beyond perturbation theory, we consider \cref{eq:Neural_Field} with a Goldstone control signal moving along $\phi(t)$, i.e., $I(x,t)=\left( c-\dot{\phi}(t)\right)U_c'(x-\phi(t))=-\dot{\phi}_1U_c'\left(x-\phi(t)\right)$,
\begin{align}
\partial_t u(x,t)=-u(x,t)+\int_{-\infty}^{\infty}\omega(x-x')\Theta(u(x',t)-\theta)\mathrm{d}x'-\dot{\phi}_1(t)\, U_c'(x-\phi(t)).
\end{align}
Assuming that the controlled solution is a TW with velocity $\dot{\phi}$, we introduce co-moving coordinates and obtain
\begin{subequations}
\begin{align}
-\dot{\phi}(t)U_c'(\xi)&=-U_c(\xi)+\int_{-\infty}^{\infty}\omega(\xi-\xi')\Theta(U_c(\xi')-\theta)\mathrm{d}\xi'-\dot{\phi}_1\,U_c'(\xi)\\
\Leftrightarrow -c\,U_c'(\xi)&=-U_c(\xi)+\int_{-\infty}^{\infty}\omega(\xi-\xi')\Theta(U_c(\xi')-\theta)\mathrm{d}\xi'.\label{eq:TW_for_proof}
\end{align}
\end{subequations}
\cref{eq:TW_for_proof} is true for the unperturbed TW solution $U_c(\xi)$ propagating with velocity $c$.
Hence, TWs with the conserved shape of the free solution, following any desired velocity protocol $\dot{\phi}(t)$ are solutions to the Goldstone-controlled NF system.

\FloatBarrier

\section{Examples on Position Control}
\label{sec:Control_app}
Within this section, we discuss generic examples of position control of TW solutions. In particular, we cover the control of traveling fronts and immobile bump solutions by means of different additive inputs in \cref{sec:subInputs} and a kernel- and threshold-mediated control of both solutions in \cref{sec:subKernThresh}.
Within the latter, we derive an expression for control by kernel modulations that is equivalent to a direct excitation of the solutions' translational mode.
\subsection{Position Control by Additive Inputs}\label{sec:subInputs}
\begin{figure}[!tb]
\center
\includegraphics[width=0.7\textwidth]{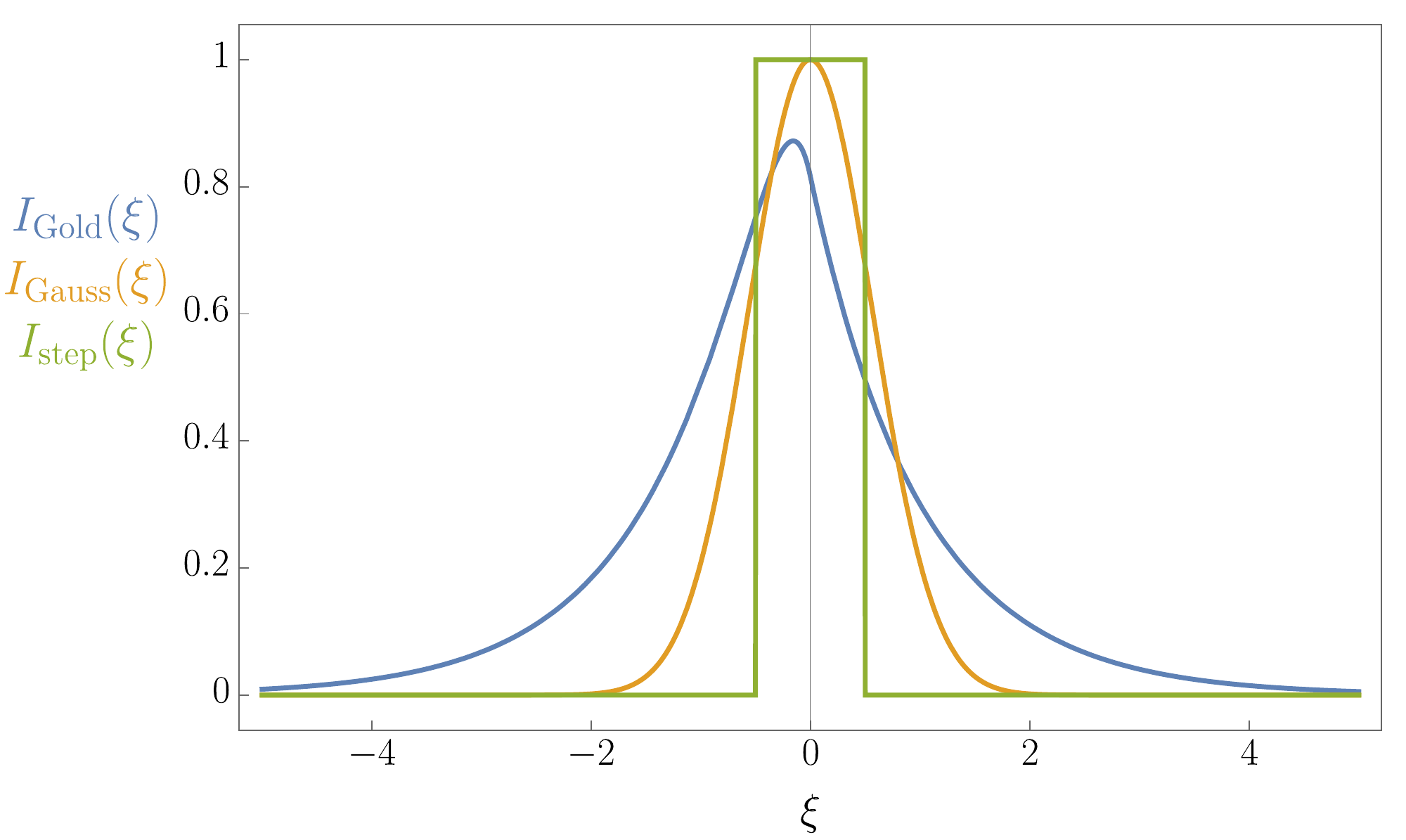}
\caption{\label{fig:inputs}Goldstone- $I_{\mathrm{Gold}}$, Gaussian- $I_{\mathrm{Gauss}}$, and step-like $I_{\mathrm{step}}$ control inputs, see \cref{eq:inputs}. The Goldstone mode shown here represents the first spatial derivative of the front solution \cref{eq:front} with $\theta=0.4$. The widths of Gaussian- and step-like inputs are $\Delta_{\mathrm{Gauss}}=0.33$ and $\Delta_{\mathrm{step}}=0.31$, respectively.}
\end{figure}

\begin{figure}[!tb]
\center
\includegraphics[width=0.7\linewidth]{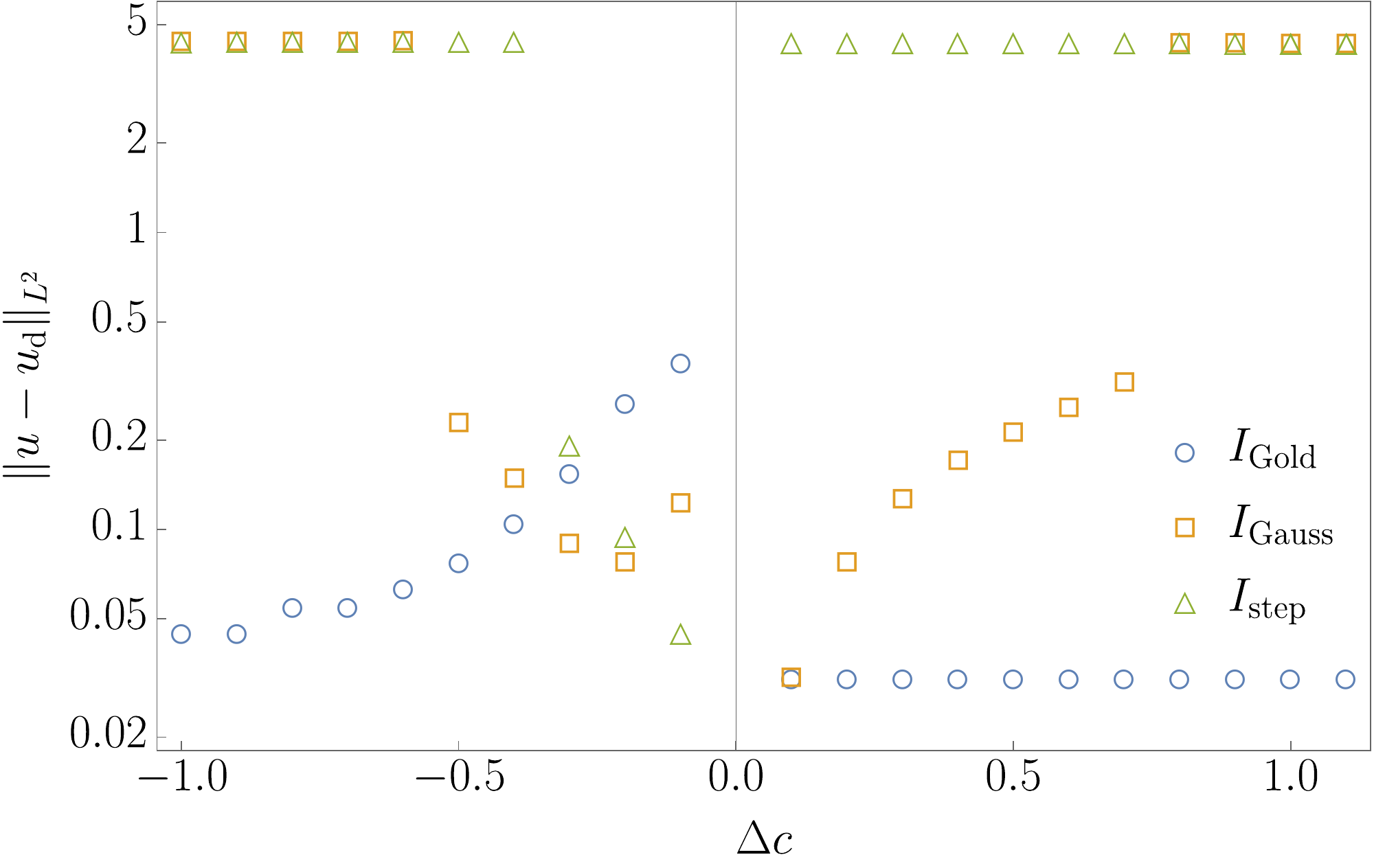}
\caption{\label{fig:input_comparison}Comparison of $L^2$-deviations between controlled and desired front solutions for different velocity changes $\Delta c$ realized by Goldstone- (circles), Gaussian- (squares), and step-control inputs (triangles), cf. \cref{eq:inputs}. The threshold $\theta=0.4$ leads to an intrinsic front velocity of $c=0.25$.}
\end{figure}

\begin{figure}[!tb]
\center
\includegraphics[width=0.7\textwidth]{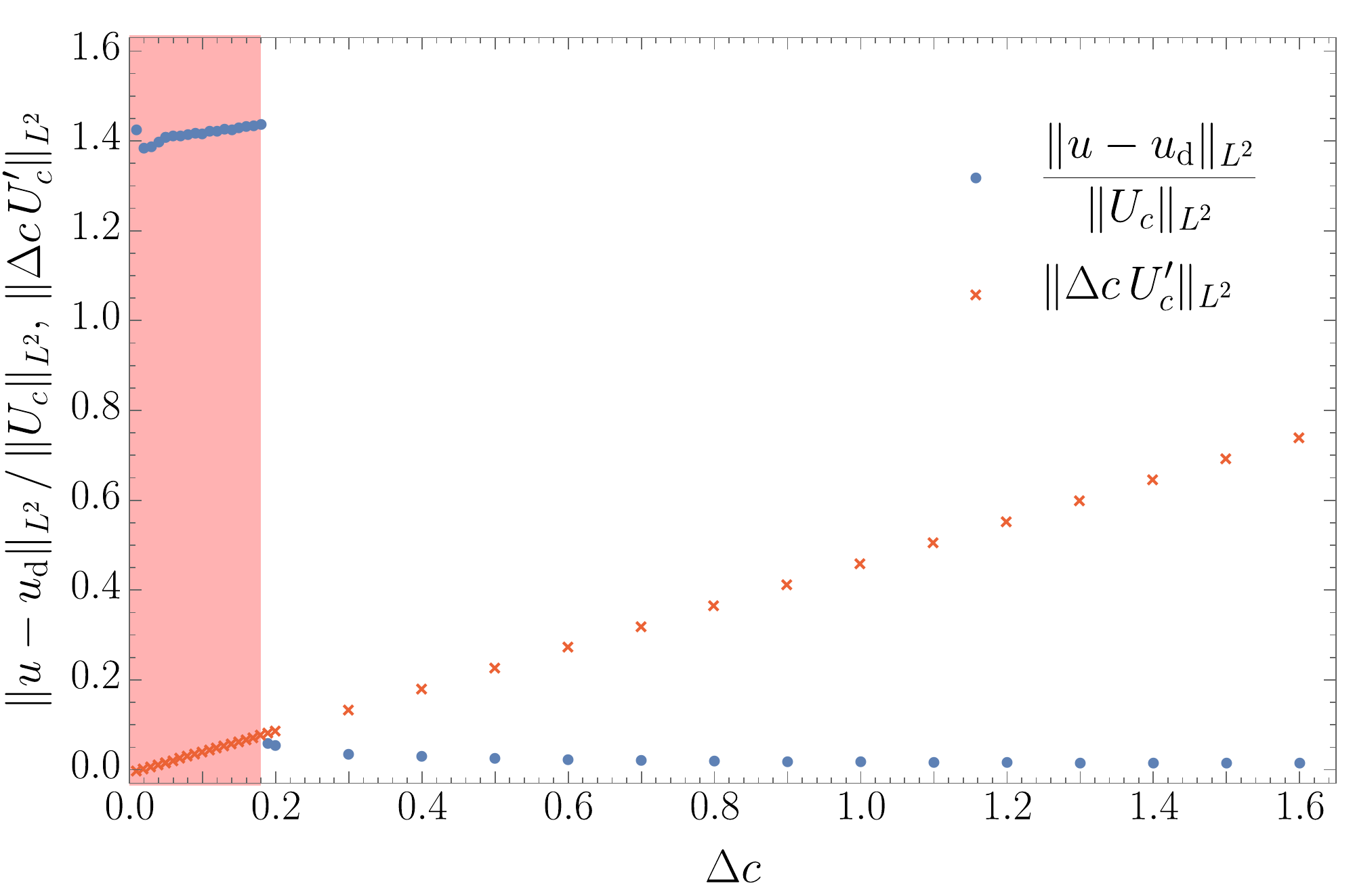}
\caption{\label{fig:bump_gold}Normalized $L^2$-deviations between a Goldstone controlled intrinsically immobile bump solution, \cref{eq:bump_sol}, and the desired state. The latter is the unperturbed solution shifted with velocity $\Delta c$. We observe a failure of control for $\Delta c \lesssim 0.18$ due to instabilities of the scheme (shaded in red). Additionally, the $L^2$-norm of the applied control signal $I(x,t)=-\Delta c\,U_c'(x-\phi(t))$ is shown.}
\end{figure}

Incorporating additional inputs $I(x,t)=\kappa_{\mathrm{ctrl}}(t)I_{\mathrm{ctrl}}(x-\phi(t))$ moving along $\phi(t)$ with amplitudes $\kappa_{\mathrm{ctrl}}(t)$ as perturbations to the NF system, the EOM \cref{eq:EOM} reduces to
\begin{align}\label{eq:EOM_front}
\dot{\phi}(t)=c-\frac{\kappa_{\mathrm{ctrl}}(t)}{K_c}\int_0^{\infty}W^{\dagger}(\xi)I_{\mathrm{ctrl}}(\xi)\mathrm{d}\xi.
\end{align}
The response functions are given by
$W^{\dagger}(\xi)=-\Theta(\xi)e^{-2\theta\xi/(1-2\theta)}$ and $W^{\dagger}(\xi)=\delta(\xi)-\delta(\xi+a)$ for front \cref{eq:front} (see orange dashed line in \cref{fig:solFront}) and immobile bump solutions \cref{eq:bump_sol}, respectively \cite{Bressloff2001,Bressloff2005}.\\
In a next step, we compare position control by direct excitation of the solution's Goldstone mode with commonly studied control signals such as step-like and Gaussian inputs \cite{Kilpatrick2012,Lu2011,Troy2007}.
The considered controls read
\begin{subequations}\label{eq:inputs}
\begin{align}
I_{\mathrm{Gold}}(x)&=U_c'(x),\\
I_{\mathrm{step}}(x)&=\Theta (x+\Delta_{\mathrm{step}}/2)-\Theta (x-\Delta_{\mathrm{step}}/2),\\
I_{\mathrm{Gauss}}(x)&=e^{-x^2/\Delta_{\mathrm{Gauss}}^2},
\end{align}
\end{subequations}
and are shown in \cref{fig:inputs}.
Besides the amplitudes $\kappa_{\mathrm{ctrl}}$, the controls incorporate free parameters $\Delta_{\mathrm{ctrl}}$ that determine the widths of the applied inputs.
We fix these quantities by maximizing the controls' impact on wave propagation, $\Delta c(t)=\dot{\phi}_1(t)$, for any given value of the input's $L^2$-norm, $\|f(x)\|_{L^2}=(\int_{-\infty}^{\infty}\lvert f(x)\rvert^2\mathrm{d}x)^{1/2}$. For a front solution propagating with $c=0.25$ ($\theta=0.4$), see \cref{eq:front}, the parameters are determined to be $\Delta_{\mathrm{step}}=0.31$ and $\Delta_{\mathrm{Gauss}}=0.33$.\\
The applicability of the considered inputs for position control of front solutions is compared in \cref{fig:input_comparison}. As a measure of success, we look at the $L^2$-deviation between the controlled and the desired state, $\|u(x)-u_{\mathrm{d}}(x)\|_{L^2}$.
Thereby, $u_{\mathrm{d}}(x)$ is defined as the unperturbed TW solution $U_c(\xi)$ shifted in space in accordance with the pre-defined velocity protocol.
Here, we choose the velocity to instantaneously jump from the unperturbed $c$ to the constant target value $c+\Delta c$,
\begin{align}
\dot{\phi}(t)=\begin{cases}
c &,t<t_0,\\
c+\Delta c &,\mbox{else.}
\end{cases}
\end{align}
In numerical simulations, the deviations are computed over an interval of 40 units of length centered around the desired wave position $\phi(t)$ with $u_{\mathrm{d}}(\phi(t),t)=\theta$, cf.\@ \cref{fig:solFront}, after a transient time.
At first glance, one notices that for Goldstone control (circles), deviations remain acceptably small over the whole range of considered velocity changes, viz., $\| u-u_d\|<0.5$. This confirms successful position control by direct excitation of the front's translational mode. We point out that shape deviations for small negative $\Delta c$ are caused by an offset between front position and control input as the proposed open-loop control scheme has a stable configuration with non-vanishing displacement, see \cref{sec:stability}.
For small to intermediate velocity changes, $\lvert\Delta c\rvert\lesssim 0.6$, the Gaussian control input (squares) yields similarly small errors and, thus, allows for successful position control as well.
Nevertheless, a Gaussian input fails to control the solution for larger changes $\lvert\Delta c\rvert\gtrsim 0.6$, mainly due to control induced shape deformations.
In cases of failure, the offset between actual and desired wave positions grows continuously and $L^2$-shape deviations are only truncated by the limited domain in which both solutions are compared.
Noteworthy, position control by a step input (triangles) fails for almost all velocity changes.
This failure can be explained considering the stability of the position control scheme with respect to offsets between the actual wave position and the location where the control signal is applied \cite{Lober2014a}. As such considerations are a key ingredient for a comprehensive treatment of position control, we included the main ideas of the method in
\cref{sec:stability}. One finds that the suggested position control scheme of traveling front solutions as \cref{eq:front} is generally unstable for a step input.
Nevertheless, the weak instability of step-like position control might be stabilized by numerical discretization effects for small negative velocity changes, $-0.3\lesssim \Delta c \leq 0$.
%
The overwhelming applicability of the Goldstone control is caused by the targeted excitation of the solution's propagation modes. Thus, within the linear approximation, no deformations of the wave profile are excited.
For TW solutions whose shape is stable against sufficiently small perturbations, the largest real-valued eigenvalue of the linear stability operator, $\lambda_0$, is zero. This eigenvalue is related to the translational invariance of the solution. If the spectral gap between $\lambda_0$ and deformation-related eigenvalues is too small, control signals can induce deformations which will not decay sufficiently fast and may eventually lead to a failure of the suggested control scheme.\\
Next, we proceed with Goldstone control of a resting bump solution. 
In \cref{fig:bump_gold}, we present the normalized $L^2$-deviation between controlled and desired state as a function of the imposed velocity change $\Delta c$. One notices large deviations, stating a failure of position control, for small velocity changes $\Delta c\lesssim 0.2$ (shaded area).
For such small values of $\Delta c$, perturbations by the control signal are small and the assumption of shape conservation holds almost perfectly. Therefore, the considerations in \cref{sec:stability} are valid and position control fails due to instabilities of the scheme. For stronger velocity changes $\Delta c \gtrsim 0.2$, the larger control amplitudes lead to a locking mechanism that outweighs small inaccuracies in applying the control that otherwise would lead to a failure of the scheme. Thus, for $\Delta c \gtrsim 0.2$, $L^2$-shape deviations are below $5\%$ of $\|U_c\|_{L^2}$. In accordance with \cref{eq:EOM_front}, the $L^2$-norm of the applied control input grows linearly with $\Delta c$.

\subsection{Position Control by Modulation of Synaptic Coupling and Neural Firing Threshold}\label{sec:subKernThresh}
Within this section, we study the influence of kernel and threshold modulations on wave propagation.
If the applied control introduces or enhances an existing break of symmetry in the system, it offers the possibility to serve as a mechanism for position control of TW solutions. 
The modulations are assumed to act equally at each position on the spatial domain. Hence, even though an open-loop scheme can be established using kernel and threshold modulations, no long term stability can be ensured. The system is expected to be marginally stable against offsets between real and desired wave position. Thus, such offsets would not necessarily lead to a total failure of control for future times. In order to compare the different means of control, we investigate whether the derived relation between desired velocity changes and necessary control amplitudes can be confirmed by numerical simulations.
If the numerically measured changes agree well with the desired ones, one is able to use the method for position control. 

\subsubsection{Modulation of synaptic coupling}
\begin{figure}[!tb]
\center
\includegraphics[width=0.7\textwidth]{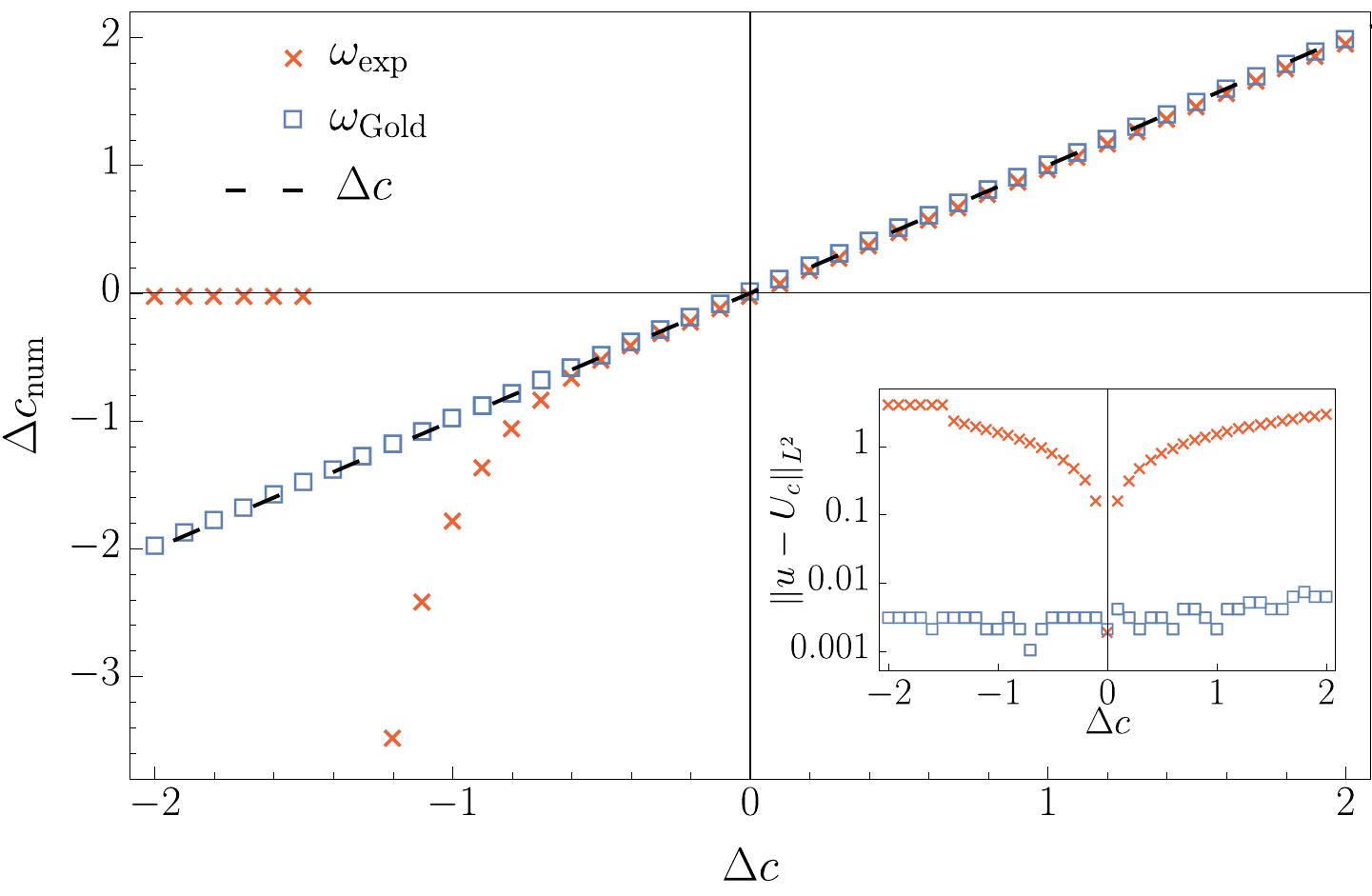}
\caption{\label{fig:kernel_front}Comparison of numerically measured velocity changes $\Delta c_{\mathrm{num}}$ caused by kernel modulations $\omega_{\mathrm{Gold}}$ and $\omega_{\mathrm{exp}}$, see \cref{eq:kernelFront1,eq:kernelFront2} for a front solution \cref{eq:front}. Only the Goldstone equivalent kernel $\omega_{\mathrm{Gold}}$ yields agreement of velocities over the whole range of desired changes $\Delta c$ while retaining shape deviations $\| u-U_c\|_{L^2}$ small. The threshold is set to $\theta=0.4$, leading to the unperturbed velocity $c=0.25$. The characteristic decay length of the exponential kernel modulation $\omega_{\mathrm{exp}}$ is set to $\varpi=1$.}
\end{figure}

For control schemes that rely on an additive input to the system, we have seen that a targeted excitation of the solution's Goldstone mode leads to an overall successful position control of TW patterns.
This is why we derive a kernel modulation, $\omega(x)\rightarrow\omega_0(x)+\omega_1(x)$, equivalent to a Goldstone input and compare it with generic modulations for which the EOM \cref{eq:EOM} can be inverted explicitly. 
For a synaptic kernel to be equivalent to a Goldstone control input, $\omega_1=\omega_{\mathrm{Gold}}$, we have the condition
\begin{align}\label{eq:equivKernCond}
\int_{-\infty}^{\infty}\omega_{\mathrm{Gold}}(x-y)f\left(U_c(y-\phi(t))\right)\mathrm{d}y=\left(c-\dot{\phi}(t)\right) U_c'(x-\phi(t)).
\end{align}
This is a direct consequence of the underlying NFE \cref{eq:Neural_Field}. In analogy to the case of a Goldstone control input, we assume that the TW follows the control protocol $\phi(t)$ while retaining the shape of the unperturbed solution.
Performing Fourier transform 
\begin{align}
\mathcal{F}\left(f(x,t)\right)\left[k,t\right]=\tilde{f}=\frac{1}{\sqrt{2\pi}}\int_{-\infty}^{\infty}f(x,t)e^{-ikx}\mathrm{d}x
\end{align}
of \cref{eq:equivKernCond} with respect to $x$ yields
\begin{align}\label{eq:omGold1}
ik\left( c-\dot{\phi}(t)\right)\,\tilde{U}_c=\sqrt{2\pi}\,\tilde{\omega}_{\mathrm{Gold}}\tilde{f}(U_c).
\end{align}
From \cref{eq:NFE_TW}, we obtain 
\begin{align}
\tilde{U}_c=\sqrt{2\pi}\,\frac{\tilde{\omega}_0\tilde{f}(U_c)}{1-ikc}
\end{align}
for the Fourier transform of stationary TW solutions.
Employing this expression in \cref{eq:omGold1}, one gains
\begin{align}\label{eq:omGold}
\tilde{\omega}_{\mathrm{Gold}}=\frac{ik\,\tilde{\omega}_0}{1-ikc}\left(c-\dot{\phi}(t)\right).
\end{align}
Hence, a kernel modulation equivalent to a Goldstone control input is given by
\begin{align}\label{eq:kernelGold}
\omega_{\mathrm{Gold}}(x,t)=\frac{c-\dot{\phi}(t)}{\sqrt{2\pi}}\int_{-\infty}^{\infty}\frac{ik\,\tilde{\omega}_0}{1-ikc}e^{ikx}\mathrm{d}k.
\end{align}
Alternatively, this expression can be rewritten in real space as
\begin{align}\label{eq:kernelGoldReal}
\omega_{\mathrm{Gold}}(x,t)=\frac{c-\dot{\phi}(t)}{c}e^{x/c}\int_{x}^{\infty}e^{-x'/c}\omega_0'(x')\mathrm{d}x'.
\end{align}
In the limit of immobile solutions $c\rightarrow 0$, the Goldstone kernel modulation $\omega_{\mathrm{Gold}}$ converges to the first spatial derivative of the free kernel $\omega_0$ with amplitude $-\Delta c$. This is in accordance with derivations made by Zhang for bumps in sheets of head direction cells \cite{Zhang1996}.
\paragraph{Position control of front solution by kernel modulations}
For the scalar NF system with coupling kernel $\omega_0(x)=0.5e^{-\lvert x\rvert}$ we can calculate a closed expression for the necessary kernel modulation. Additionally, we are able to solve the EOM for other control kernels and, therefore, we can compare the targeted excitation of the solution's translational mode with other, more arbitrary modulations.
On the basis of \cref{eq:kernelGold}, the Goldstone kernel is given by
\begin{align}\label{eq:kernelFront2}
\omega_{\mathrm{Gold}}(x)=\frac{c-\dot{\phi}(t)}{2\pi}\int_{-\infty}^{\infty}\frac{ik}{1-ikc}\frac{1}{1+k^2}e^{ikx}\mathrm{d}k,
\end{align}
which can be solved using residue theorem, leading to
\begin{align}
\omega_{\mathrm{Gold}}(x)=-\left(c-\dot{\phi}(t)\right)\begin{cases}
\frac{e^{-x}}{2\left(c+1\right)} & , x\geq 0,\\
\frac{e^x}{2\left(c-1\right)}-\frac{e^{x/c}}{c^2-1} & , x<0.
\end{cases}
\end{align}
As a modulation to compare this kernel with, we choose a generic asymmetric exponential kernel
\begin{align}\label{eq:kernelFront1}
\omega_{\mathrm{exp}}(x)=\begin{cases}
\kappa e^{-\varpi x}  &,\ x\geq 0,\\
0 &\mbox{, else.}
\end{cases}
\end{align}
Its amplitude $\kappa$ can be calculated in dependence on the desired velocity changes $-\left(c-\dot{\phi}(t)\right)=\Delta c$ using \cref{eq:EOM},
\begin{align}
\Delta c=\dot{\phi}_1(t)=-\frac{\kappa}{K_c\varpi\left(\varpi+1/c\right)}.
\end{align}
With $c=1/(2\theta)-1$ and $K_c=\theta\left(2\theta-1\right)$, we obtain the kernel amplitude
\begin{align}
\kappa=\varpi\theta\left(\varpi+2\theta(1-\varpi)\right)\Delta c
\end{align}
for traveling fronts in the scalar NFE.
In \cref{fig:kernel_front}, the impact $\Delta c_{\mathrm{num}}$ of both control kernels $\omega_{\mathrm{exp}}(x)$ and $\omega_{\mathrm{Gold}}(x)$ on the propagation velocity is shown for different desired velocity changes $\Delta c$.
The threshold parameter is set to $\theta=0.4$, which leads to a front solution \cref{eq:front} with unperturbed velocity $c=0.25$. Further, the length scale of the control kernel $\omega_{\mathrm{exp}}$ is kept fixed to $\varpi=1$.\\
We observe an overwhelming agreement between desired and resulting propagation velocities, viz., $\Delta c_{\mathrm{num}}\approx \Delta c$, for kernel modulations $\omega_{\mathrm{Gold}}$ equivalent to a Goldstone input. Thereby, the $L^2$- shape deviations $\|u-u_{\mathrm{d}}\|_{L^2}$ from the unperturbed front solution, $u_{\mathrm{d}}(x,t)=U_c(x-\phi(t))$, remain remarkably small over the whole range of velocity changes $\Delta c\in\left[-2,2\right]$. In comparison, for the exponential kernel modulation $\omega_{\mathrm{exp}}$, errors grow larger for negative velocity changes, $\Delta c\lesssim -0.5$ as not only the translational mode is excited by the kernel modulation. The resulting shape deformations, see inset in \cref{fig:kernel_front}, can lead to deviations from the predicted velocity. For stronger decelerating protocols, the control kernel outweighs the unperturbed synaptic coupling so that almost no more excitation occurs and the front solution vanishes for $\kappa <0.5$.
\paragraph{Control of immobile bump solution by kernel modulations}
Next, we investigate the applicability of kernel mediated position control on immobile bump solutions. For these solutions, the derived expression for $\omega_{\mathrm{Gold}}$ becomes equal to the first spatial derivative of the free coupling kernel $\omega_0$,
\begin{align}\label{eq:gold_kernel_bump}
\omega_{\mathrm{Gold}}(x)=-\frac{\Delta c}{\sqrt{2\pi}}\int_{-\infty}^{\infty}ik\,\tilde{\omega}_0e^{ikx}\mathrm{d}k=-\Delta c\,\omega_0'(x).
\end{align}
For simplicity, we again assumed constant velocity changes $\Delta c$.
Exemplarily, we chose the free kernel of the bump solution to be given by \cref{eq:kernel_bump} and compare the results of the Goldstone equivalent kernel control with another representative asymmetric exponential control kernel, namely
\begin{align}\label{eq:asym_kernel}
\omega_{\mathrm{exp}}(x)=\begin{cases}
\frac{\kappa}{2}e^{-\varpi x} &, x\geq 0,\\
-\frac{\kappa}{2}e^{\varpi x} &, \mbox{else.}
\end{cases}
\end{align}
The EOM \cref{eq:EOM} with any kernel modulations reads
\begin{align}
\dot{\phi}_1(\tau)=-\frac{1}{K_c}\int_{-\infty}^{\infty}\left(\delta(\xi)-\delta(\xi+a)\right)\int_{-\infty}^{\infty}\omega_1(\xi-\xi')\Theta(U_c(\xi')-\theta)\mathrm{d}\xi'\mathrm{d}\xi,
\end{align}
for immobile bump solutions \cref{eq:bump_sol} and yields the expression
\begin{align}\label{eq:expKernCtrlBump}
\kappa=-\varpi K_c\left(e^{-\varpi a}-1\right)^{-1}\Delta c,
\end{align}
for $\omega_1(x)=\omega_{\mathrm{exp}}(x)$ with constant $K_c=2U_c'(0)$. This relation allows to calculate a necessary control in terms of the amplitude $\kappa$ of the kernel modulation \cref{eq:asym_kernel} for a pre-defined velocity change $\Delta c$.
\begin{figure}
\center
\includegraphics[width=0.8\textwidth]{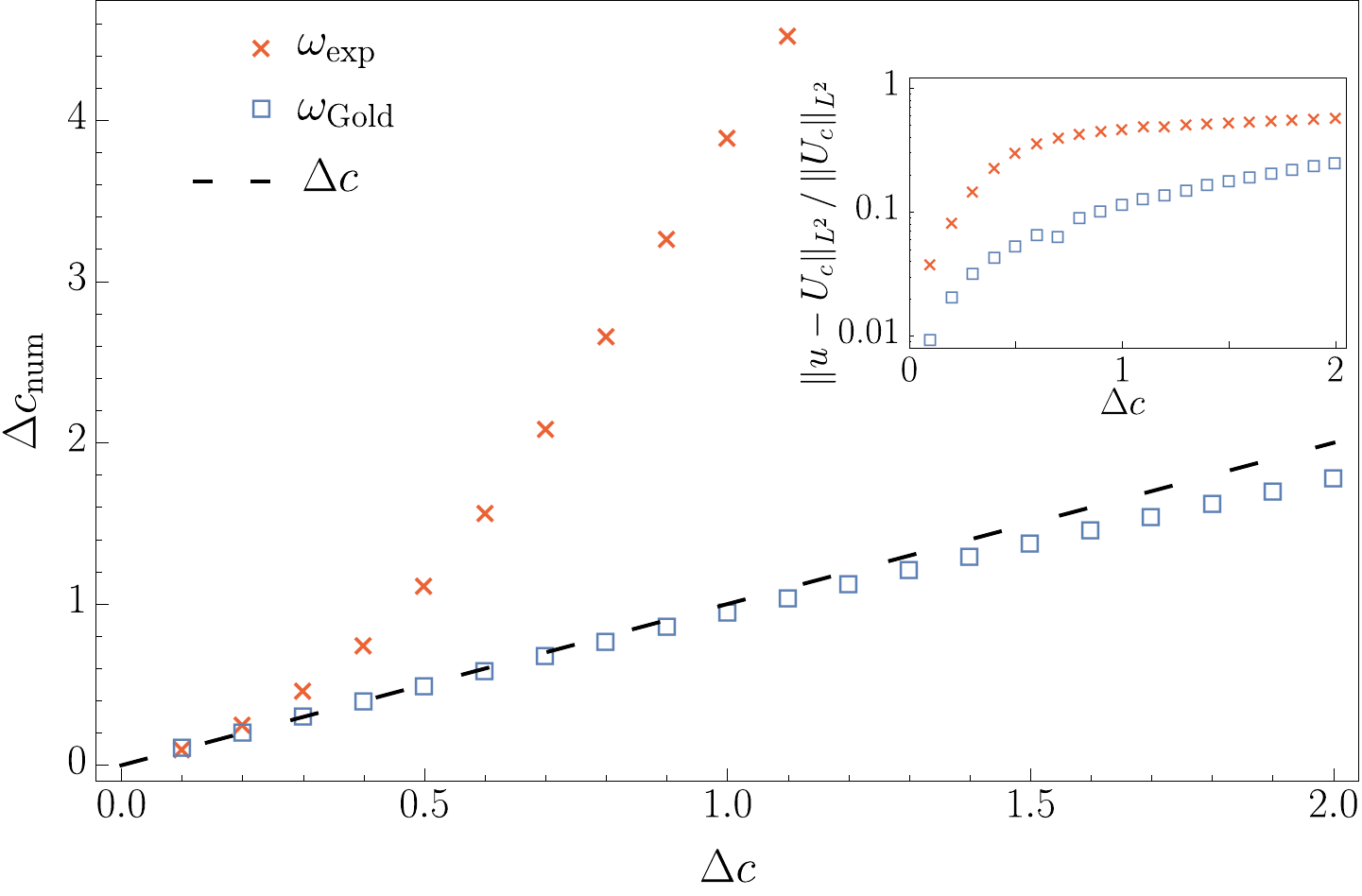}
\caption{\label{fig:bump_kernel}Velocity changes $\Delta c_{\mathrm{num}}$ of an immobile bump solution, cf.\@ \cref{eq:bump_sol} with free synaptic kernel \cref{eq:kernel_bump}, affected by Goldstone equivalent $\omega_{\mathrm{Gold}}$ (blue squares) and exponentially decaying $\omega_{\mathrm{exp}}$ (orange crosses) kernel modulations, see \cref{eq:gold_kernel_bump,eq:asym_kernel}, respectively.
The resulting velocities of control by $\omega_{\mathrm{Gold}}$ agree with the desired values $\Delta c$ (dashed line) over a large range of target velocities. 
The $L^2$-shape deviations of the Goldstone kernel controlled solution in units of the $L^2$ norm of the unperturbed bump (inset) are one order of magnitude below the ones for the exponential control kernel. The parameters are $\theta=0.077$, and $\varpi=1$.}
\end{figure}\\
In \cref{fig:bump_kernel}, the desired velocity changes $\Delta c$ are compared with results $\Delta c_{\mathrm{num}}$ of corresponding numerical simulations for $\varpi=1$. The inset depicts the numerically measured shape deformations $\|u-U_c\|_{L^2}/\|U_c\|_{L^2}$ between the controlled and the free solution in units of the latter's $L^2$-norm. Thereby, offsets between actual and desired wave position were compensated by shifting both solutions to a reference position identifying the right-hand sides' threshold crossing points of both bump solutions.\\
For comparably weak control kernels, $\Delta c\lesssim 0.2$, velocities of controlled bump solutions agree well with the desired changes for both examined kernels. Similarly to the case of traveling front solutions, shape deviations $\| u-U_c\|_{L^2}$ grow larger for the exponential kernel (inset in \cref{fig:bump_kernel}) leading to much higher velocities than desired. With increasing modulation amplitudes, the induced shape deformations cause a failure of the approximations made during perturbation analysis. Hence, linear perturbation theory \cref{eq:expKernCtrlBump} is no longer valid and resulting velocities differ noticeably from desired ones.
In contrast, the Goldstone control kernel yields solutions with velocities close to the target values $\Delta c$ while maintaining shape deformations below $20\%$ even for very large velocity changes $\Delta c\gtrsim 1.5$.
Again, we point out that even though position control by arbitrary kernel modulations is applicable, the resulting shape deformations are nowhere as small as for the Goldstone control kernel. This, in turn, leads to less predictable velocity changes for non-Goldstone kernels. In consequence, position control by spatio-temporal kernel modulations can be strongly simplified using Goldstone mode equivalent synaptic weights.
For stationary, immobile patterns it would be necessary to measure the activity distribution and to extract the firing function by single neuron experiments. Subsequently, the synaptic kernel $\omega_0(x)$ can be reconstructed by minimizing an appropriate error functional \cite{Zhang1996}.

\subsubsection{Control of immobile bump solutions by threshold modulations}\label{sec:subThresh}

\begin{figure}[!tb]
\center
\includegraphics[width=0.7\textwidth]{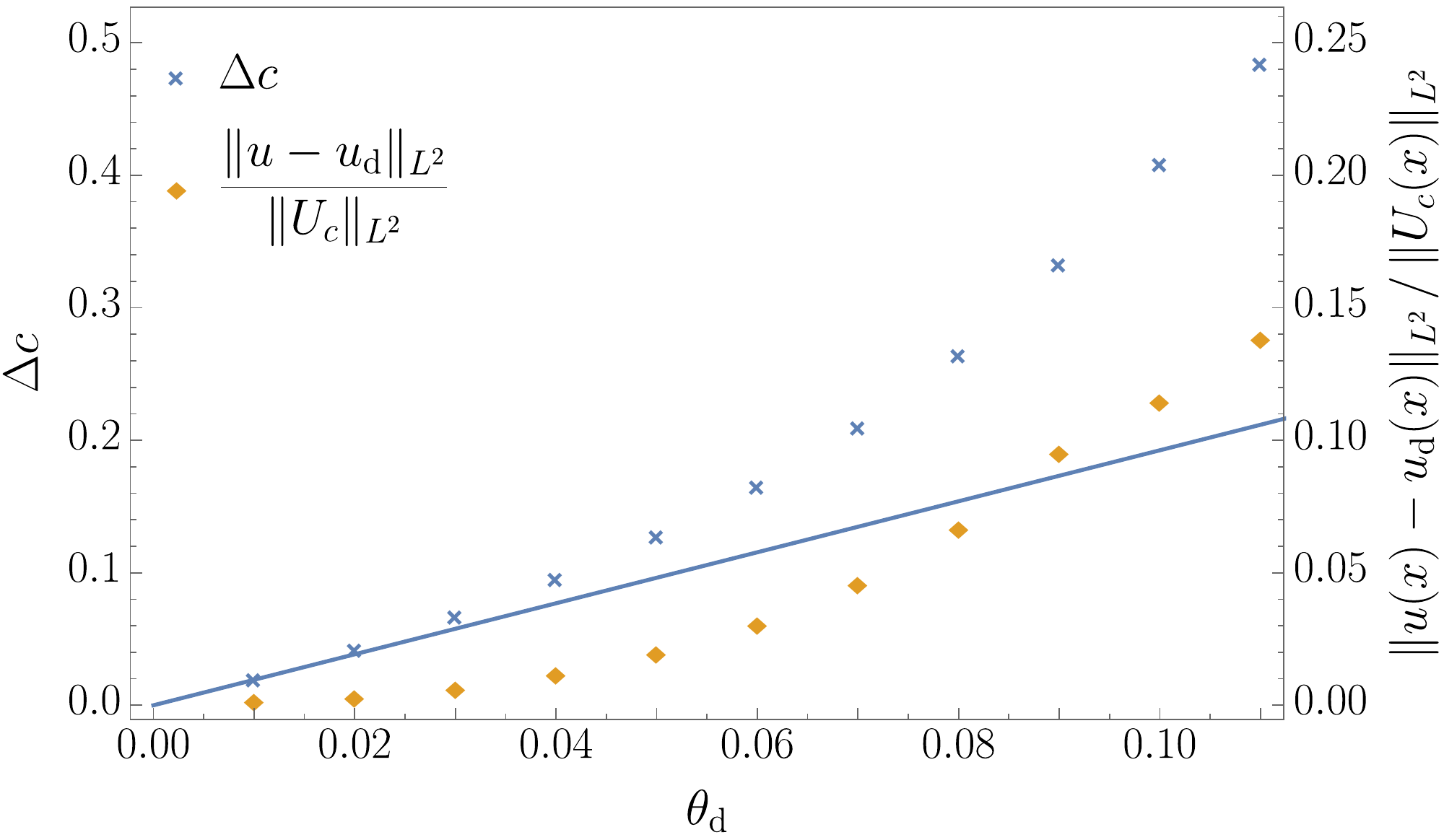}
\caption{\label{fig:bump_thresh}Numerically measured (crosses) and analytically predicted (solid line) velocity changes $\Delta c$ of an immobile bump solution \cref{eq:bump_sol} caused by threshold modulations according to \cref{eq:asym_thresh}. The synaptic coupling is given by \cref{eq:kernel_bump} and the unperturbed threshold value is set to $\theta_0=0.077$. The resulting shape deviations $\|u-u_{\mathrm{d}}\|_{L^2}$ (diamonds) grow monotonously with increasing control amplitude $\theta_{\mathrm{d}}$.}
\end{figure}

As a last example, we cover control by spatio-temporal modulations of the neuronal firing threshold $\theta=\theta_0+\theta_1(x,t)$. In particular, we exemplarily choose the control
\begin{align}\label{eq:asym_thresh}
\theta_1(x,t)=\theta_1(\xi)=\begin{cases}
\theta_{\mathrm{d}} &,-\delta-a\leq\xi\leq\delta-a,\\
-\theta_{\mathrm{d}} &,-\delta\leq\xi\leq\delta,\\
0 &, \mbox{else}
\end{cases}
\end{align}
to be given by a bi-modal step-like increase and decrease of the neural firing threshold at the two flanks of the bump solution \cref{eq:bump_sol}, respectively. There, its propagation behavior is most sensitive to perturbations as can be seen by the contributions of the bump solution's response function $W^{\dagger}(\xi)=\delta(\xi)-\delta(\xi+a)$ with bump width $a$. To assure long time stability of the scheme, we implement the control by shifting the threshold modulation following the movement of the controlled bump solution $\xi=x-\phi(t)$. The width of the applied modulation is set to $2\delta=0.8$.
Incorporating this control term \cref{eq:asym_thresh}, the EOM \cref{eq:EOM} reduces to
\begin{align}
\dot{\phi}_1(\tau)&=\frac{1}{K_c}\int_{-\infty}^{\infty}\left(\delta(\xi)-\delta(\xi+a)\right)\int_{-\infty}^{\infty}\omega(\xi-\xi')\left(\frac{\delta(\xi')}{\lvert U_c'(0)\rvert}+\frac{\delta(\xi'+a)}{U_c'(-a)}\right)\theta_1(\xi')\mathrm{d}\xi'\mathrm{d}\xi\nonumber\\
&=\frac{-2\theta_d}{K_c\lvert U_c'(0)\rvert}\left(\omega(0)-\omega(-a)\right)
\end{align}
and, straightaway, we end up with the relation
\begin{align}\label{eq:thresh_rel}
\theta_{\mathrm{d}}=-\frac{K_c\lvert U_c'(0)\rvert}{2\left(\omega(0)-\omega(a)\right)}\Delta c.
\end{align}
Equation \cref{eq:thresh_rel} gives an expression for the necessary amplitudes $\theta_{\mathrm{d}}$ of threshold modulation that cause a change of propagation velocity of the controlled bump solution by $\Delta c$.
In \cref{fig:bump_thresh}, the numerically measured velocities for different values $\theta_{\mathrm{d}}$ are compared with the theoretical predictions. 
We observe a good agreement as long as perturbations do not become too large, viz., $\theta_{\mathrm{d}}\lesssim 0.04$.
The controlled solution's shape deviations $\|u(x)-u_{\mathrm{d}}(x)\|_{L^2}$ remain below $5\%$ of $\|U_c\|_{L^2}$ as long as control amplitudes $\theta_{\mathrm{d}}\lesssim0.07$. Nevertheless, they are already large enough to cause noticeable deviations between predicted and measured propagation velocities.
This again emphasizes the disadvantages of controls that do not solely excite the solution's translational mode. Only the Goldstone control, either mediated by an additive input or a corresponding kernel modulation leads to the desired wave propagation with high accuracy over a broad range of velocity changes.
\FloatBarrier

\section{Conclusion}
 \label{sec:Conclusion}
We have treated control of activity regions in large populations of synaptically coupled neurons. In particular, we have taken advantage of neural field equations to describe the dynamics within neural tissue and controlled the spatial position of front and bump solutions in them. By means of singular perturbation analysis, we have presented an equation of motion for traveling wave solutions under perturbations such as external inputs, modified synaptic footprints, and spatial modulations of the neural firing threshold. 
In a next step, we have posed the inverse question and utilized the considered perturbations to control the position of patterns of neural activity. By solving the derived equation of motion for the amplitudes of the perturbations, we have been able to compute the necessary control for any given velocity protocol.
For arbitrary means of control, the knowledge of the model equations and the solutions' response functions is mandatory for the inversion. Consequently, one is generally limited to a few examples for which explicit control signals can be calculated. In order to overcome this limitation, we have utilized an additive control input that is proportional to the translational mode of the TW solution to be controlled.
%
For such a control signal, one can always invert the equation of motion and, therefore, obtain an explicit expression for the control amplitude. 
The proposed position control is designed as an open-loop control for which one has to measure the unperturbed solution's profile as well as its propagation velocity once with sufficient accuracy. All aspects that are relevant for the so-called Goldstone control are encoded within these measures.
This makes the Goldstone control scheme practical for various applications in which the details of the underlying neural field system are potentially unknown. To justify the presented control methods, we have explicitly carried out position control of traveling front solutions. In detail, we have compared commonly studied additive inputs such as Gaussian- and step-like signals with the suggested Goldstone control.
Besides the fact that one can always derive an explicit expression for the control amplitudes for a given desired velocity change, the Goldstone control does not induce shape deformations of the controlled solutions.
This leads to a superior applicability in comparison with other inputs.
Nevertheless, we have observed instabilities for open-loop position control of spatially symmetric bump solutions. On the basis of the derived equation of motion, conditions for the stability of the control scheme with respect to offsets between control and actual wave position have been applied. Thereby, a satisfying agreement between the analytically predicted stable offsets and the ones measured in numerical simulations has been observed.
If the control exhibits stability, one can set up the Goldstone position control as an open-loop scheme. This increases the control's applicability to cases where a monitoring of the scheme's progress is challenging. Further, the Goldstone control is spatially localized at the position of the traveling wave and, therefore, less invasive than other control methods that for example involve inputs on a global scale. It allows for an instantaneous adjustment of the solutions' propagation velocity and,
moreover, it was shown for reaction-diffusion systems, that the Goldstone control is close to an optimal control signal with appropriate regularization conditions \cite{Ryll2016}.\\
As an alternative to direct inputs to the neural field system, we have investigated position control by asymmetric synaptic coupling kernels and spatial modulations of the local firing threshold.
In particular, we have derived an expression for a spatio-temporal kernel modulation that is equivalent to the Goldstone input and, thus, allows for explicit position control without evoking an excitation of deformation modes.
In contrast to generic kernel modulations, the Goldstone kernel yields velocity changes close to the desired values while retaining the shape of the unperturbed solution over broad ranges of velocities.
Finally, we want to emphasize that most of the techniques discussed in this work can be extended to more involved systems such as multi-component neural field models in one or two spatial dimensions.\\

\FloatBarrier

\appendix
\section{Stability of open-loop position control of TW solutions to the neural field equations}
\label{sec:stability}
\begin{figure}[!tb]
\center
\includegraphics[width=0.65\linewidth]{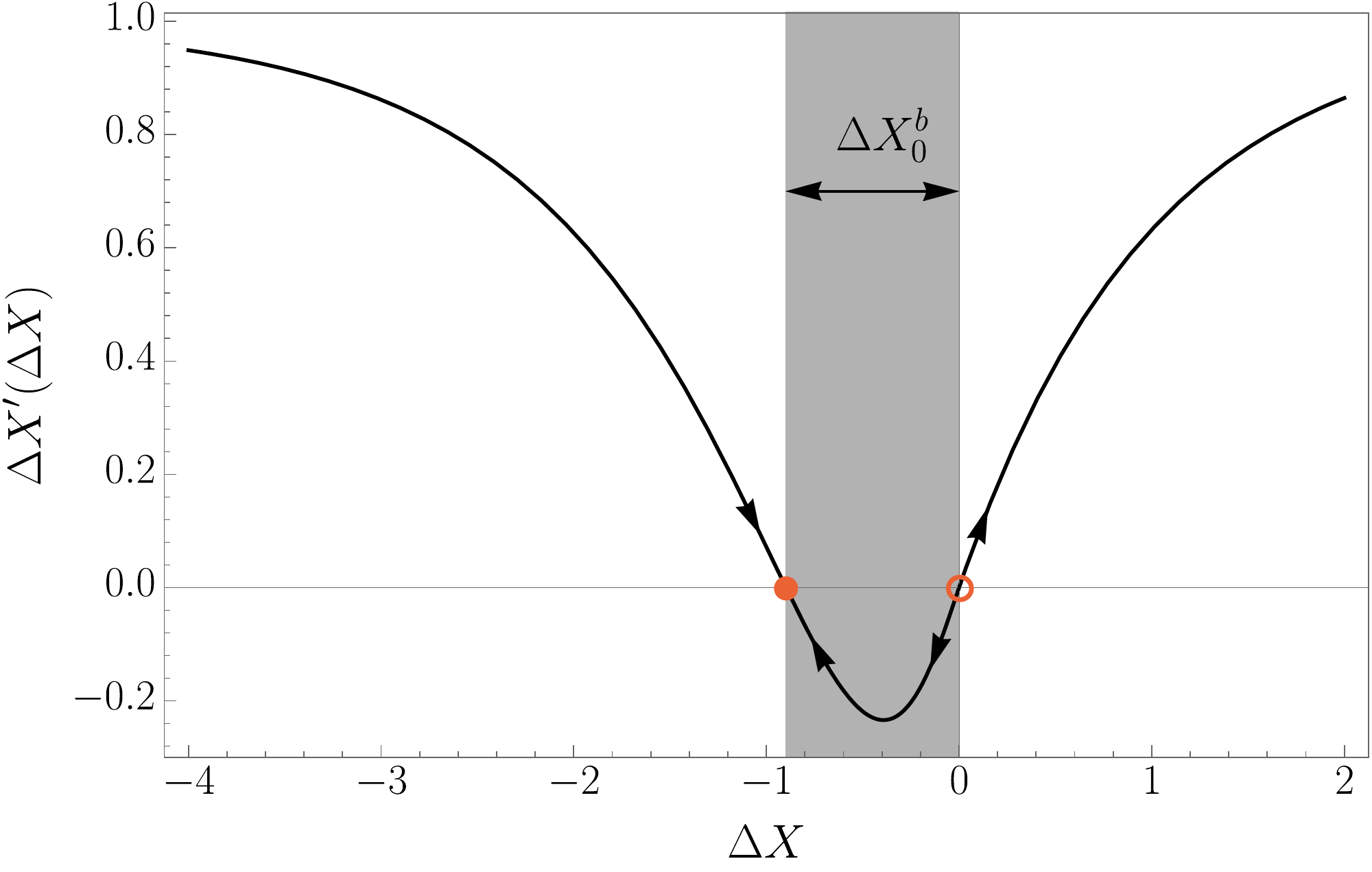}
\caption{\label{fig:stab_Gold1}Right-hand side of \cref{eq:stab_dyn} for a front solution \cref{eq:front} with Goldstone control input and threshold $\theta=0.4$. The offset variable $\Delta X$ possesses two fixed points, $\Delta X_0^a<0$ and $\Delta X_0^b=0$, whose stability can be investigated on the basis of the linear growth rate $\lambda_1(\Delta X)\rvert_{\Delta X=\Delta X_0}$ around the fixed points, see \cref{eq:lambda_stab}.}
\end{figure}

\begin{figure}[!tb]
\center
\includegraphics[width=\textwidth]{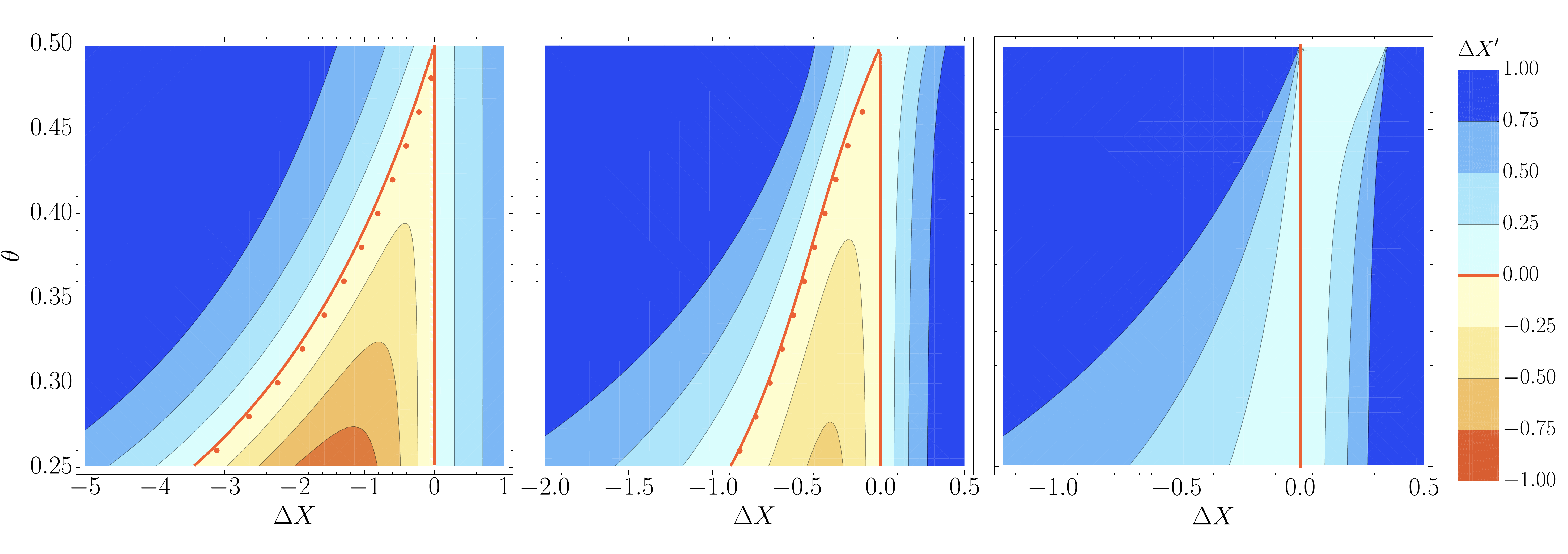}
\caption{\label{fig:stab_GoldVSStep}Right-hand side of  \cref{eq:stab_dyn} for a front solution \cref{eq:front} with Goldstone- (left panel), Gaussian- (center panel), and step- (right panel) control input. The fixed points are determined by the condition $\Delta X'(z) = 0$ for different thresholds $\theta$. In numerical simulations (orange dots), we set the desired velocity change to $\Delta c = -0.05$.}
\end{figure}

For completeness, we outline the main steps of stability analysis of the input based control scheme. In particular, we apply a method introduced by L\"ober \cite{Lober2014a} for TW solutions to reaction-diffusion systems to the neural field equations.\\
So far, we have derived an EOM \cref{eq:EOM} for the position of traveling wave solutions under perturbations in \cref{sec:Control}. By solving this equation for a chosen perturbation, one can calculate the control that leads to a propagation following a desired velocity protocol. 
Nevertheless, the successful inversion of the EOM does not imply stability of the control scheme as can be seen in the examples of position control of traveling front solutions by step inputs as well as for control of immobile bump solutions in \cref{sec:subInputs}.
Stability of control methods is a key ingredient for establishing an open-loop scheme which shall not require any further examination of the controlled solution but solely rely on preliminary measurements.
In consequence, throughout the course of control, the pre-computed control signal is always applied as if the controlled TW solution followed perfectly the desired protocol.\\
On the basis of EOM \cref{eq:EOM} we investigate the dynamics of the offset $\Delta X(ct-X(t))=\phi(t)-X(t)$ between desired $X(t)$ and actual wave position $\phi(t)$.
For simplicity, we restrict ourselves to the case of additive control inputs $\kappa_{\mathrm{ctrl}}(t)I_{\mathrm{ctrl}}(\xi)$, yielding
\begin{equation}\label{eq:stab_dyn}
\Delta X'(z)=1+\frac{\kappa_{\mathrm{ctrl}}}{\Delta c\,K_c}\int_{-\infty}^{\infty}W^{\dagger}(\xi)I_{\mathrm{ctrl}}(\xi+\Delta X(z))\mathrm{d}\xi,
\end{equation}
with solution's response function $W^{\dagger}(\xi)$, desired velocity change $\Delta c$, and new coordinate $z(t)=ct-X(t)$.
We consider the control as successful if $\Delta X(z(t))$ remains bounded for all times $t$.
In order to predict the long time evolution of the offset variable $\Delta X(z(t))$, we search for fixed points $\Delta X_0$ of the system. These are given by the condition $\Delta X'(\Delta X_0)=0$ and their linear stability is governed by the growth rate
\begin{align}\label{eq:lambda_stab}
\lambda=\frac{\kappa_{\mathrm{ctrl}}}{\Delta c\,K_c}\int_{-\infty}^{\infty}W^{\dagger}(\xi)\left[\partial_{\xi'}I_{\mathrm{ctrl}}(\xi')\right]_{\xi'=\xi+\Delta X_0}\mathrm{d}\xi.
\end{align}
For Goldstone control of the front solution, see \cref{eq:front}, $\kappa_{\mathrm{ctrl}}I_{\mathrm{ctrl}}(\xi)=-\Delta c\,U_c'(\xi)$, with $U_c'(x)=-U_c''(x)\,\forall\, x>0$ and $W^{\dagger}(\xi)=-\Theta(\xi)e^{-2\theta\xi/(1-2\theta)}$, \cref{eq:lambda_stab} gives
\begin{align}
\lambda_0=\frac{1}{K_c}\int_0^{\infty}e^{-2\theta\xi/(1-2\theta)}U_c''(\xi)\mathrm{d}\xi=1
\end{align}
at the fixed point $\Delta X_0=0$.
We stress, that the evolution of the deviation $\Delta X(z)$ in \cref{eq:stab_dyn} incorporates the coordinate $z(t)$ which does not necessarily grow in time $t$.
Hence, the linear stability of fixed points is not solely determined by the sign of $\lambda$ but it also depends on the sign of the desired velocity change: If $\Delta c>0$, $z$ decreases over time $t$ as $X(t)>ct$. Thus, for $\lambda<0$ and $\lambda>0$ a fixed point is unstable and stable, respectively. This behavior is inverted for negative desired velocity changes $\Delta c<0$.
In \cref{fig:stab_Gold1}, the right-hand side of \cref{eq:stab_dyn} is illustrated for the case of Goldstone control of a front solution as given in \cref{eq:front}.
We observe that two fixed points of the offset variable $\Delta X$ coexist. As expected, one of them is located at zero offset $\Delta X_0^a=0$. For decelerating protocols, this is an unstable fixed point, whereas $\Delta X_0^b<0$ is stable. Consequently, control schemes with initial offsets $\Delta X_{\mathrm{init}} < 0$ exhibit long time stability for a decelerating Goldstone control.
Such a behavior appears sensible if one exemplarily thinks of a decelerating control signal ahead of the TW solution. The TW will eventually approach the slower control until the control acts properly on the solution and $\Delta X$ approaches the stable fixed point.\\
In \cref{fig:stab_GoldVSStep}, the threshold dependent locations of the fixed points are shown for front solutions \cref{eq:front} with Goldstone- (left panel), Gaussian- (center panel), and step-control inputs (right panel).
While two fixed points of the control scheme coexist for the Goldstone as well as a Gaussian control signal (left and center panel, respectively), there is only a single one for step-like signals (right panel). For the exemplarily chosen decelerating control protocol ($\Delta c = -0.05$) in \cref{fig:stab_GoldVSStep}, the fixed point $\Delta X_0=0$ is unstable for Goldstone and Gaussian control which is in agreement with the analytical predictions (solid orange lines). Moreover, we emphasize that the analytically derived values of $\Delta X_0^b$ are validated by numerical simulations. The corresponding orange dots indicate the values of the stable fixed point against which the setup converges to for initial offsets $\Delta X_{\mathrm{init}}^1=-0.2$ as well as $\Delta X_{\mathrm{init}}^2=-3.7$.
Hence, we find that there exist regions of stability of the control scheme for Goldstone- and Gaussian inputs for front solutions \cref{eq:front}. In contrast, the configuration is intrinsically unstable for a step-like control input leading to a long time divergence of the offset value $\Delta X$.\\
The considerations presented above, can be analogously executed for position control of immobile bump solutions \cref{eq:bump_sol}. With a Goldstone control input, $I_{\mathrm{ctrl}}(\xi)=U_c'(\xi)$, the dynamic equation for the offset reads
\begin{align}
\Delta X'(z)=1-\frac{1}{K_c}\left(U_c'(\Delta X(z))-U_c'(\Delta X(z)-a)\right),
\end{align}
where we use the bump's response function $W^{\dagger}(\xi)=\delta(\xi)-\delta(\xi+a)$ with bump width $a$. The fixed point condition reduces to
\begin{align}
2U_c'(0)=U_c'(\Delta X)-U_c'(\Delta X-a),
\end{align}
with $K_c=2U_c'(0)$.
Due to the symmetry of the solution, the position control scheme exhibits only a single fixed point \cite{Lober2014a}.
As the linear growth rate $\lambda_1$ is discontinuous at $\Delta X_0=0$ and
\begin{align}
\lim_{\Delta X\nearrow 0}\lambda_1 <0, \qquad \lim_{\Delta X\searrow 0}\lambda_1 >0
\end{align}
we identify the fixed point to be a saddle. This does not allow for a long-time stable configuration for position control, confirming the failure of the scheme in numerical simulations as shown in \cref{fig:bump_gold}.

\FloatBarrier

\section*{Acknowledgments}
We gratefully acknowledge financial support by the German Science Foundation DFG through collaborative research center SFB 910.
\newpage
\bibliographystyle{siamplain}
\bibliography{paperBib}
\end{document}